\documentclass[sigconf, letterpaper]{acmart}
\usepackage{graphicx}
\usepackage[utf8]{inputenc}
\usepackage[english]{babel}
\usepackage{url}
\usepackage{color}
\usepackage{textcomp}
\usepackage{svg}
\usepackage{multirow}
\usepackage{pgfplots}
\usepackage{listings}
\DeclareMathOperator*{\argmin}{arg\,min}
\pgfplotsset{compat=newest}
\usepgfplotslibrary{groupplots}
\usepgfplotslibrary{dateplot}
\usepackage[colorinlistoftodos]{todonotes}
\presetkeys{todonotes}{inline}{}

\AtBeginDocument{%
    \providecommand\BibTeX{{%
\normalfont{}B\kern-0.5em{\scshape i\kern-0.25em b}\kern-0.8em\TeX}}}

\newcommand{\ssep}{\ensuremath{\,|\,}}

\newcommand{\toolname}{{WAF-A-MoLE}}
\newcommand{\sqligot}{SQLiGoT}
\newcommand{\wafbrain}{WAF-Brain}
\newcommand{\modsecurity}{ModSecurity}
\newcommand{\tokenbased}{Token-based}

\newcommand{\scikit}{\emph{scikit-learn}}
\newcommand{\sota}{state of the art}

\begin{document}
\title{\toolname: Evading Web Application Firewalls through Adversarial Machine Learning}

\author{Luca Demetrio}
\email{luca.demetrio@dibris.unige.it}
\affiliation{\institution{Università di Genova}}
\author{Andrea Valenza}
\email{andrea.valenza@dibris.unige.it}
\affiliation{\institution{Università di Genova}}
\author{Gabriele Costa}
\email{gabriele.costa@imtlucca.it}
\affiliation{\institution{IMT School for Advanced Studies Lucca}}
\author{Giovanni Lagorio}
\email{giovanni.lagorio@unige.it}
\affiliation{\institution{Università di Genova}}


\settopmatter{printacmref=false} 
\renewcommand\footnotetextcopyrightpermission[1]{} 
\pagestyle{plain} 
\copyrightyear{2020}
\acmYear{2020}
\setcopyright{acmcopyright}
\acmConference[SAC '20]{The 35th ACM/SIGAPP Symposium on Applied Computing}{March 30-April 3, 2020}{Brno, Czech Republic}
\acmBooktitle{The 35th ACM/SIGAPP Symposium on Applied Computing (SAC '20), March 30-April 3, 2020, Brno, Czech Republic}
\acmPrice{15.00}
\acmDOI{10.1145/3341105.3373962}
\acmISBN{978-1-4503-6866-7/20/03}

\begin{abstract}
    Web Application Firewalls are widely used in production environments to mitigate security threats like SQL injections.
    Many industrial products rely on signature-based techniques, but machine learning approaches
    are becoming more and more popular.
    The main goal of an adversary is to craft \emph{semantically} malicious payloads to bypass the \emph{syntactic} analysis performed by a WAF.

    In this paper, we present \toolname{}, a tool that models the presence of an adversary.
    This tool leverages on a set of mutation operators that alter the syntax of a payload without affecting the original semantics.
    We evaluate the performance of the tool against existing WAFs, that we trained using our publicly available SQL query dataset.
    We show that \toolname{} bypasses all the considered machine learning based WAFs.
\end{abstract}

\maketitle

\section{Introduction}\label{sec:intro}

Most security breaches occur due to the exploitation of some vulnerabilities.
Ideally, the best way to improve the security of a system is to detect all its vulnerabilities and patch them.
Unfortunately, this is rarely feasible due to the extreme complexity of real systems and high costs of a thorough assessment.
In many contexts, payloads arriving from the Internet are the primary threat, with the attacker using them to discover and exploit some existing vulnerabilities.
Thus, protecting a system against malicious payloads is crucial.
Common protection mechanisms include input filtering, sanitization, and other domain-specific techniques, e.g.,~\emph{prepared statements}.
Implementing effective input policies is non trivial and, sometimes, even infeasible (e.g., when a system must be integrated in many heterogeneous contexts).

For this reason, mitigation solutions are often put in place.
For instance, \emph{Intrusion Detection Systems} (IDS) aim to detect suspicious activities.
Clearly, these mechanisms have no effect on existing vulnerabilities that silently persist in the system.
However, when IDSs can precisely identify intrusion attempts, they significantly reduce the overall damage.
The very core of any IDS is its detection algorithm: the overall effectiveness only depends on whether it can discriminate between harmful and harmless packets/flows.

Web Application Firewalls (WAFs) are a prominent family of IDS, widely adopted~\cite{Gartner17waf} to protect ICT infrastructures.
Their detection algorithm applies to HTTP requests, where they look for possible exploitation patterns, e.g., payloads carrying a SQL injection.
Since WAFs work at application-level, they have to deal with highly expressive languages such as SQL and HTML.
Clearly, this exacerbates the detection problem.

\begin{figure}
    \begin{lstlisting}[
           language=SQL,
           basicstyle=\ttfamily,
           commentstyle=\color{gray},
           showspaces=false,
           showstringspaces=false,
           numbers=left,
           breakindent=1em,
           breaklines=true,
           xleftmargin=1.8em
        ]
admin' OR 1=1#
admin' OR 0X1=1 or 0x726!=0x726 OR 0x1Dd not IN/*(seleCt 0X0)>c^Bj>N]*/ ((SeLeCT 476),(SELECT (SElEct 477)),0X1de) oR 8308 noT lIkE  8308\x0c AnD truE OR 'FZ6/q' LiKE 'fz6/qI' anD TRUE anD '>U' != '>uz'#t'%'03;Nd
    \end{lstlisting}
    \caption{Two semantically equivalent payloads.}
    \label{fig:sqli-example}
\end{figure}

To clarify this aspect, consider a classical SQL injection scenario where the attacker crafts a malicious payload $x$ such that the query \texttt{SELECT * FROM users WHERE name=\textquotesingle$x$\textquotesingle\ AND pw=\textquotesingle$y$\textquotesingle} always succeeds (independently from $y$).
Figure~\ref{fig:sqli-example} shows two instances of such a payload.
Notice that the two payloads are semantically equivalent.
As a matter of fact, both reduce the above query to
\texttt{SELECT * FROM users WHERE name=\textquotesingle{}admin\textquotesingle{} OR $\top$ \#$\ldots$} where $\top$ is a tautology and $\ldots$ is a trail of commented characters.
Ideally a WAF should reject both these payloads.
However, when classification is based on a mere syntactical analysis, this might not happen.
Hence, the goal of an attacker amounts to looking for some malicious payload that is undetected by the WAF.
We present a technique to effectively and efficiently generate such malicious payloads, that bypass ML-based WAF.
Our approach starts from a target malicious payload that the WAF correctly detects.
Then, by iteratively applying a set of mutation operators, we generate new payloads.
Since mutation operators are semantics-preserving, the new payloads are equivalent from the point of view of the adversary.
However, they gradually reduce the confidence of the WAF classification algorithm.
Eventually, this process converges to a payload classified below the rejection threshold.
To evaluate the effectiveness of our methodology we implemented a working prototype, called \toolname{}.
Then we applied \toolname{} to different ML-based WAFs, and evaluated their robustness against our technique.

\noindent
\textbf{Contributions of the paper.}
The main contributions of this work are summarized as follows: $(i)$ we develop a tool for producing adversarial examples against WAFs by leveraging on a set of syntactical mutations, $(ii)$ we produce a dataset of both sane and injection queries, $(iii)$ and we review the state of the art of machine learning SQL injection classifiers and we bypass them using \toolname{}.

\section{Preliminaries}
\label{sec:preliminary}


Web Application Firewalls (WAFs) are commonly used to prevent application-level exploits of web applications.
Intuitively, the idea is that a WAF can detect and drop dangerous HTTP requests to mitigate potential vulnerabilities of web applications.
The most common detection mechanisms include \emph{signature-based matching} and \emph{classification via machine learning}.

Signature-based WAFs identify a payload according to a list of rules, typically written by some developers or maintained by a community.
For instance, rules can be encoded through some policy specification language that defines the syntax of legal/illegal payloads.
Nowadays, the signature-based approach is widely used and, perhaps, the most popular signature-based WAF is ModSecurity\footnote{\url{https://modsecurity.org}}.

However, recently the machine learning-based approach has received increasing attention.
For instance, both FortiNet\footnote{\url{https://www.fortinet.com/blog/business-and-technology/fortiweb-release-6-0--ai-based-machine-learning-for-advanced-thr.html}} and PaloAlto\footnote{\url{https://www.paloaltonetworks.com/detection-response}} include ML-based detection in their WAF products, since
ML can overcome some limitations of signature-based WAFs, i.e., the extreme complexity of developing a list of syntactic rules that precisely characterizes malicious payloads.
Since ML WAFs are trained on existing legal and illegal payloads, their configuration is almost automatic.


\emph{Adversarial machine learning} (AML)~\cite{barreno2006can,huang2011adversarial} studies the threats posed by an attacker aiming to mislead machine learning algorithms.
More specifically, here we are interested in \emph{evasion attacks}, where the adversary crafts malicious payloads that are wrongly classified by the victim learning algorithm.
The adversarial strategy varies with the target ML algorithm.
Many existing systems have been shown to be vulnerable and several authors, e.g.~\cite{biggio13-ecml,goodfellow2014explaining,papernot2016limitations,carlini2017adversarial}, proposed techniques for systematically generating malicious samples.
Intuitively, the crafting process works by introducing a \emph{semantics-preserving} perturbation in the payload, that interferes with the classification algorithms.
Notice that, often, a formal semantics of the classification domain is not available, e.g., it is informally provided through an oracle such as a human classifier.
The objective of the adversary may be written as a constrained minimization problem
$x^* = \argmin_{x, \mathcal{C}(x)} \mathcal{D}(f(x), c_t)$,
where $f$ is the victim classifier, $c_t$ is the desired class the adversary wants to reach, $\mathcal{D}$ is a distance function,
and $\mathcal{C}(x)$ represents all the constraints that cannot be violated during the search for adversarial examples.
Since we consider binary classifiers, we can rewrite our problem as $x^* = \argmin_{x, \mathcal{C}(x)} f(x)$, where the output of $f$ is bounded between $0$ and $1$, and we are interested in reaching the benign class represented by $0$.

\begin{table*}[t]
    \begin{tabular}{l l l}
        \toprule
        Operator & Short definition & Example \\
        \midrule
        Case Swapping & $CS(\ldots a \ldots B \ldots) \rightarrow \ldots A \ldots b \ldots$ & $CS($\verb|admin' OR 1=1#|$) \rightarrow $ \verb|ADmIn' oR 1=1#| \\
        Whitespace Substitution & $WS(\ldots k_1 k_2 \ldots) \rightarrow \ldots k_1$ \textvisiblespace{} $k_2 \ldots$ & $WS($\verb|admin' OR 1=1#|$) \rightarrow$ \verb|admin'\n OR \t 1=1#|\\
        Comment Injection & $CI(\ldots k_1 k_2 \ldots) \rightarrow \ldots k_1$\verb|/**/|$k_2\ldots$ & $CI($\verb|admin' OR 1=1#|$) \rightarrow$ \verb|admin'/**/OR 1=1#| \\
        Comment Rewriting & $CR(\ldots$\verb|/*|$s_0$\verb|*/|$\ldots$\verb|#|$s_1) \rightarrow \ldots$\verb|/*|$s'_0$\verb|*/|$\ldots$\verb|#|$s'_1$ &
        $CR($\verb|admin'/**/OR 1=1#|$) \rightarrow $ \verb|admin'/*abc*/OR 1=1#xyz| \\
        Integer Encoding &  $IE(\ldots n \ldots) \rightarrow \ldots 0x[n]_{16}$ & $IE($\verb|admin' OR 1=1#|$) \rightarrow $ \verb|admin' OR 0x1=1#| \\
        Operator Swapping & $OS(\ldots \oplus \ldots) \rightarrow \ldots \boxplus \ldots$ (with $\oplus \equiv \boxplus$) & $OS($\verb|admin' OR 1=1#|$) \rightarrow $ \verb|admin' OR 1 LIKE 1#| \\
        Logical Invariant & $LI(\ldots e \ldots) \rightarrow \ldots e$ \verb|AND| $\top \ldots$ & $LI($\verb|admin' OR 1=1#|$) \rightarrow $ \verb|admin' OR 1=1 AND 2<>3#| \\
        \bottomrule
    \end{tabular}
    \caption{List of mutation operators.}\label{tab:mutation-set}
\end{table*}
\section{Overview of \toolname{}}
\label{sec:overview}

\begin{figure}[t]
    \includegraphics[width=\columnwidth]{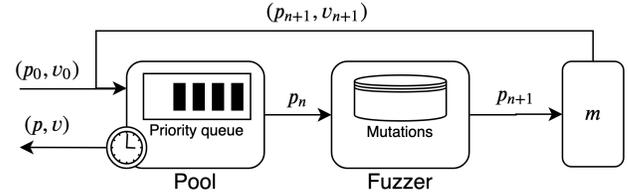}
    \caption{An outline of the mutational fuzz testing approach.}
    \label{fig:tool-fig}
\end{figure}

Our methodology belongs to the class of \emph{guided mutational fuzz testing} approaches~\cite{mutationalfuzzing, fuzzingbook2019:MutationFuzzer}.
Briefly, the idea is to start from a failing test, that gets repeatedly transformed through the random application of some predefined mutation operators.
The modified tests, called \emph{mutants}, are then executed, compared (according to some performance metric) and ordered.
Then, the process is iterated on the tests that performed better until a successful test is found.
Clearly, this approach requires both a comparison criterion and a set of mutation operators.
These are typically application-dependent.
Figure~\ref{fig:tool-fig} schematically depicts this approach.

\subsection{Algorithm description}\label{sec:algorithm}

In our context a test is a SQL injection and its execution amounts to submitting it to the target WAF\@.
The comparison is based on the confidence value generated by the detection algorithm of the WAF\@.
The payload pool is the data structure containing the SQL injection candidates to be mutated during the next round.
Below we describe in more detail the set of mutation operators and the payload pool.

\let\origthelstnumber\thelstnumber
\makeatletter
\newcommand*\Suppressnumber{%
    \lst@AddToHook{OnNewLine}{%
        \let\thelstnumber\relax%
        \advance\c@lstnumber-\@ne\relax%
    }%
}

\newcommand*\Reactivatenumber{%
    \lst@AddToHook{OnNewLine}{%
        \let\thelstnumber\origthelstnumber%
    \advance\c@lstnumber\@ne\relax}%
}

\makeatother

\begin{figure}
    \Suppressnumber
    \begin{lstlisting}[language=Pascal,mathescape,escapeinside=||, morekeywords={input,classify,enqueue,foreach,end,mutate,output,head,exit},numbers=left,xleftmargin=2em]
  input: Model $m$, Payload $p_0$, Threshold $t$
  output: head($Q$)
|\Reactivatenumber|
  $Q$ := create_priority_queue()
  $v$ := classify($m$, $p_0$)
  enqueue($Q$, $p_0$, $v$)
  while $v > t$
    $p$ := mutate(head($Q$))
    $v$ := classify($m$, $p$)
    enqueue($Q$, $p$, $v$)
    \end{lstlisting}
    \caption{Core algorithm of \toolname{}.}\label{fig:algorithm}
\end{figure}

A pseudo code implementation of the core algorithm of \toolname{} is shown in Figure~\ref{fig:algorithm}.
The algorithm takes the learning model $m : \mathcal{X} \rightarrow [0,1]$, where $\mathcal{X}$ is the feature space, an initial payload $p_0$ and a threshold $t$, i.e., a confidence value under which a payload is considered harmless.
\toolname{} implements the payload pool (see Section~\ref{sec:payload}) as a priority queue $Q$ (line 1).
The payloads in $Q$ are prioritized according to the confidence value returned by the classification algorithm, namely \textbf{classify}, associated to $m$.
The classification algorithm assigns to each payload an $x\in\mathcal{X}$, by extracting a feature vector, and computes $m(x)$.

Initially, $Q$ only contains $p_0$ (lines 2--3).
The main loop (lines 4--7) behaves as follows.
The head element of $Q$, i.e., the payload having the lowest confidence score, is extracted and mutated (line 5), by applying a set of mutation operators (see Section~\ref{sec:mutation}). The obtained payload, $p$, is finally classified (line 6) and en-queued (lines 7).
The termination of the algorithm occurs when a $p$ receives a score less or equal to the threshold $t$ (line 4).

\subsection{Mutation operators}\label{sec:mutation}

A mutation operator is a function that changes the syntax of a payload so that the semantics of the injected queries is preserved.

Below we describe the considered mutation operators.

\noindent
\textbf{CS.} The \emph{Case Swapping} operator randomly changes the capitalization of the keywords in a query (e.g., \verb|Select| to \verb|sELecT|).
        Since SQL is case insensitive, the semantics of the query is not affected.

\noindent
\textbf{WS.} \emph{Whitespace Substitution} relies on the equivalence between several alternative characters that only act as separators (whitespaces) between the query tokens.
        For instance, whitespaces include \verb|\n| (line feed), \verb|\r| (carriage return) and \verb|\t| (horizontal tab).
        Each of these characters can be replaced by an arbitrary, non-empty sequence of the others without altering the semantics of the query.

\noindent
\textbf{IC.} Inline comments (\verb|/*...*/|) can be arbitrarily inserted between the tokens of a query.
        Since comments are not interpreted, they are semantics preserving.
        The \emph{Comment Injection} operator randomly adds inline comments between the tokens.

\noindent
\textbf{CR.} Following the above reasoning, the \emph{Comment Rewriting} operator randomly modifies the content of a comment.

\noindent
\textbf{IE.} The \emph{Integer Encoding} operator modifies the representation of numerical constants.
        This includes alternative base representations, e.g., from decimal to hexadecimal, as well as statement nesting, e.g., \verb|(SELECT 42)| is equivalent to \verb|42|.

\noindent
\textbf{OS.} Some operators can be replaced by others that behave in the same way.
        For instance, the behavior of \verb|=| (equality check) can be simulated by \verb|LIKE| (pattern matching).
        We call this mutation \emph{Operator Swapping}.

\noindent
\textbf{LI.} A \emph{Logical Invariant} operator modifies a boolean expression by \emph{adding opaque predicates}.\footnote{That is, heuristically generated true and false expressions to be combined in conjunction and disjunction (respectively) with the payload clauses.}


Table~\ref{tab:mutation-set} provides a compact list, including a short, mnemonic definition, of the operators described above.

\subsection{Mutation tree}\label{sec:payload}

\begin{figure*}[t]
    \includegraphics[width=0.8\textwidth]{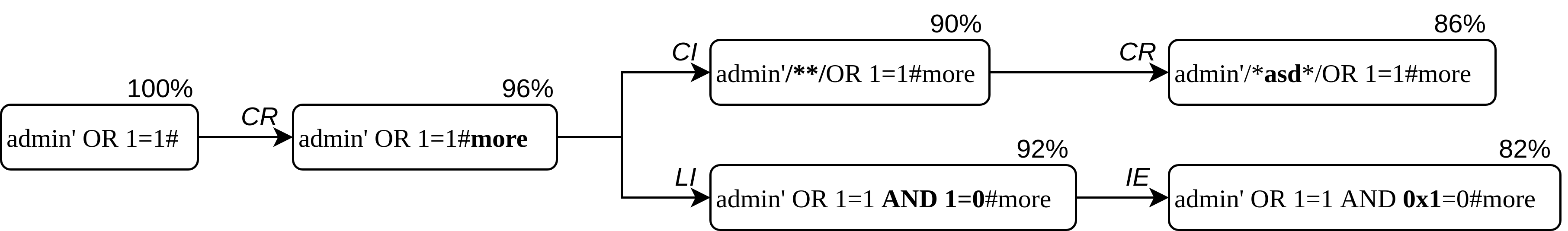}
    \caption{A possible mutation tree of an initial payload.}\label{fig:payload-population}
\end{figure*}

The priority queue of Figure~\ref{fig:algorithm} contains a sequential representation of mutation tree.
Starting from a root element, i.e., the initial payload ($p_0$ in Figure~\ref{fig:algorithm}), a mutation tree contains elements obtained through the application of some mutation operator.
A possible instance of a mutation tree is shown in Figure~\ref{fig:payload-population}.
Each edge is labeled with an identifier of the applied mutation operator.
Also, each node is labeled with a possible classification value (in percentage).
The corresponding queue is given by the sequence of the nodes in the mutation tree ordered by the associated classification value.

After applying a mutation (actually after a full \emph{mutation round}, see Section~\ref{sec:efficiency}), the payload is evaluated and added to the priority queue, along with information
about the payload that generated it.
Keeping all individuals in the initial population helps avoiding local minima:
when a payload is unable to create better payloads, the algorithm tries to backtrack on
old payloads to create a new branch on the mutation tree.

\subsection{Efficiency}\label{sec:efficiency}

The main bottleneck of our algorithm is the classification step.
Indeed, the classification of a payload requires the extraction of a vector of features.
Although a WAF classifier is efficient, the feature extraction process may require non-negligible string parsing operations (see Section~\ref{sec:models}).
For example, the procedure carried out by a token-based classifier (see Section~\ref{sec:our-wafs} for details) requires non-trivial computation to parse the SQL query language (being context free).
Instead, all the mutation operators described in Section~\ref{sec:mutation} rely on efficient string parsing, based on regular expressions.

We mitigate this issue by following a \emph{mutation preemption} strategy,
i.e., we create a \emph{mutation round} where multiple payloads are generated at once.
All these mutated payloads are stored for the classification.
Then we run all the classification steps in parallel and we discharge the mutants that increase the classification value of their parent.
In this way we take advantage of the parallelization support of modern CPUs.

For memory efficiency, we only enqueue a mutated payload if it improves the classification value of its parent.
In this way we mitigate the potential, exponential blow-up of the mutation tree (see Section~\ref{sec:payload}).
On the negative side, each branch of the mutation tree only evolves monotonically which might result in the algorithm stagnating on local minima.
However, our experiments show that this does not prevent our algorithm from finding an injectable payload (see Section~\ref{sec:attacks}).

\newcommand{\benignData}{20,000}
\newcommand{\injData}{20,000}
\newcommand{\phiBenignData}{768}
\newcommand{\phiInjData}{7,963}
\newcommand{\sqligotDBenignData}{3216}
\newcommand{\sqligotDInjData}{12,659}
\newcommand{\sqligotUBenignData}{3268}
\newcommand{\sqligotUInjData}{12,682}
\newcommand{\RFestimators}{25}

\section{WAF training and benchmarking}\label{sec:models}

Our technique applies to an input model representing a \emph{well-trained} WAF, i.e., a WAF that effectively detects malicious payloads.
Ideally, to generate a payload that bypasses a deployed WAF, the input algorithm should rely on the same detection model.
In the case of ML-based WAF, the model is the result of the training process over a sample dataset, while for signature-based WAFs the model is the set of all the collected signatures that are used as a comparison for future input data.

Unfortunately, it is very common that neither the detection model nor the training dataset are publicly available.
Reasonably, this happens because the WAF manufacturers (correctly) consider such knowledge an advantage for the adversary.
Remarkably, this also happens for the research prototypes.\footnote{All maintainers of the WAFs considered in this work were contacted, but no one provided their datasets.}
Thus, we had to create a training dataset and configure the classification algorithms.
The following sections describe the issues we faced during this process and how we solved them.

\subsection{Dataset}\label{sec:dataset}

To the best of our knowledge, no dataset of benign SQL queries is publicly available.
The main reason is probably that the notion of ``benign'' is application-dependent and no universal definition exists.
On the other hand, there are many malicious payloads, that one can extract from existing penetration testing tools such as \emph{sqlmap}\footnote{\url{https://github.com/sqlmapproject/sqlmap}} and \emph{OWASP ZAP}\footnote{\url{https://www.owasp.org/index.php/OWASP_Zed_Attack_Proxy_Project}}.
We consider the payloads generated by these tools, as any WAF should be trained on well-known attacks.

We built our dataset through an automatic procedure\footnote{The dataset is available at \url{https://github.com/blindusername/wafamole-dataset}}.
In particular, we used \emph{randgen}\footnote{https://github.com/MariaDB/randgen} to generate the queries.
Starting from a grammar $G$, the tool returns a set of queries that belong to the language denoted by $G$.
Noticeably, queries generated by \emph{randgen} also include actual values, e.g., table and column names, referring to a given existing database.
Thus, the queries in the dataset can be submitted and evaluated against a real target.

To create our labeled dataset, we assume that SQL queries are always created by the application when a user submits a payload, either benign or malicious.
To simulate this behavior, we generate a single initial grammar that supports multiple query types.
Then, we provide different dictionaries of values for each terminal symbol (i.e., $t$, $f$, $v$) that represents a possible value of a particular column inside the database.

The query grammar is the following.

    \begin{tabular}{l c l}
        $Q$ & ::= & $S \ssep U \ssep D \ssep I$ \\
        $S$ & ::= & \textbf{SELECT} $(\bar{f} \ssep *)$ \textbf{FROM} $t$ \textbf{WHERE} $e$ [\textbf{LIMIT} $\bar{v}$]\\
        $U$ & ::= & \textbf{UPDATE} $t$ \textbf{SET} $f = v$ \textbf{WHERE} $e$ [\textbf{LIMIT} $\bar{v}$] \\
        $D$ &  ::= & \textbf{DELETE FROM} $t$ \textbf{WHERE} $e$ [\textbf{LIMIT} $\bar{v}$] \\
                \end{tabular}
    
    \begin{tabular}{l c l}
        $I$ &  ::= & \textbf{INSERT INTO} $t$ ($\bar{f}$) \textbf{VALUES} ($\bar{v}$) \\
        $e$ & ::= & $f \gtreqless v \ssep f$ \textbf{LIKE} $s \ssep e$ \textbf{AND} $e' \ssep e$ \textbf{OR} $e'$  \\
    \end{tabular}
    \medskip

Briefly, the queries $Q$ can be \emph{select} $S$, \emph{update} $U$, \emph{delete} $D$ or \emph{insert} $I$.
The syntax of each query is standard, only notice that $S$, $U$ and $D$ may optionally (square brackets) terminate with a \textbf{LIMIT} clause.
The queries operate on several parameter types, including fields $f$, tables $t$, values $v$, strings $s$ and boolean expressions $e,e'$.
Finally, we use $\bar{\cdot}$ to denote a vector, i.e., a finite, comma-separated list of elements.
The actual values for $t$ and $f$ are taken from an actual target database (this feature is provided by \emph{randgen}).
For $v$, we use different values depending on the type of query we want to generate.
For the benign queries, we generate payloads with a random generator, a dictionary of nations,
a dictionary of values which are compatible with the field type to simulate a real application payload.
For example, in a database containing people names we use English first and last names.
We are interested in the structure of the query, hence these values for the payload are suitable for our analysis.

As mentioned above, the malicious values are generated by \emph{sqlmap} and \emph{OWASP ZAP}.

\subsection{Classification algorithms}\label{sec:our-wafs}

Below we describe the classification algorithms that we used for our experiments.
In particular, we consider different techniques, built on three feature extraction methods: characters, token and graph based.

\paragraph{Character-based features}
\emph{\wafbrain}\footnote{\url{https://github.com/BBVA/waf-brain}} is based on a recurrent-neural network.
The network divides the input query in blocks of exactly five consecutive characters.
Its goal is to predict the sixth character of the sequence based on the previous five.
If the prediction is correct, the block of characters is more likely to be part of a malicious payload.
This process is repeated for every block of five characters forming the target query.

The neural network of \wafbrain{} is structured as follows. The input layer is a Gated Recurrent Unit (GRU)~\cite{cho2014learning} made of five neurons, followed by two fully-connected layers, i.e., a dropout layer followed by another fully connected layer.
Finally, \wafbrain{} computes the average of all the prediction errors over the input query and scores it as malicious if the result is above a fixed threshold chosen a priori by the user.
Since the threshold is not given by the classifier itself,
as all the other details of the training and cross-validation phases, we set it to 0.5, which is the standard threshold for classification tasks.

\paragraph{Token-based features}
The token-based classifiers represent input queries as histograms of symbols, namely tokens.
A token is a portion of the input that corresponds to some syntactic group, e.g., a keyword, comparison operators or literal values.

We took inspiration from the review written by Komiya et al.~\cite{komiya2011classification} and Joshi et al~\cite{joshi2014sql} and we developed a tokenizer for producing the features vector to be used by these models.
On top of that, we implemented different models: $(i)$ a Naive Bayes (NB) classifier, $(ii)$ a random forest (RF) classifier with an ensemble of \RFestimators{} trees, $(iii)$ a linear SVM (L-SVM), and a gaussian SVM (G-SVM).
We trained them using a 5-fold cross-validation with \benignData{} sane queries and \injData{} injections, and we used 15\% of the queries for the validation set.
To this extent, we coded our experiment using \scikit{}~\cite{scikit-learn}, which is a Python library containing already implemented machine learning algorithms.
\begin{table}[]
    \begin{tabular}{llrrrr}

        \toprule
                        & & \textbf{C} & \boldmath$\gamma$  & \boldmath$avg(A)$ & \textbf{\boldmath$\sigma$}\\
						\midrule
		\multirow{4}{*}{Token-based}
        & Naive Bayes     & /      & /     & 54.2\% & 1.0\% \\
        & Random Forest   & /      & /     & 87.3\% & 0.7\% \\
        & Linear SVM      & 19.30  & /     & 80.5\% & 1.4\% \\
        & Gaussian SVM    & 278.25   & 0.013 & 93.1\% & 0.9\% \\
		\midrule
		\multirow{4}{*}{\sqligot{}}		
        & Dir. Prop.   &4.64&0.26& 99.85\% &0.07\%\\
        & Undir. Prop. &2.15&0.71& 99.10\%  & 0.2\%\\
        & Dir. Unprop. &2.15&0.26& 99.74\% &0.1\%\\
        & Undir. Unprop.&2.15&0.26& 98.89\% & 0.2\%\\
        \bottomrule
    \end{tabular}
    \caption{Training phase results.}
    \label{table:token_based}
\end{table}
After the feature extraction phase, the number of samples dropped to \phiBenignData{} benign and \phiInjData{} injection queries.
The tokenization method is basically an aggregation method: only a subset of all symbols are taken into account.
The dataset is unbalanced, as the variety of sane queries is outnumbered by the variety of SQL injections.
To address this issue, we set up \scikit{} accordingly, by using a loss function that takes into account the class imbalance~\cite{brodersen2010balanced}.
Table~\ref{table:token_based} shows the results of the training phase where $(i)$ $C$ is the regularization parameter~\cite{tikhonov1943stability} that controls the stability of the solution, $(ii)$ $\gamma$ is the kernel parameter (only for the gaussian SVM)~\cite{aizerman1964theoretical,hofmann2008kernel}, and $(iii)$ $avg(A)$ and $\sigma$ are the average and standard deviation of the accuracy computed during the cross-validation phase over the validation set.
\paragraph{Graph-based features}
Kar et al.~\cite{kar2016sqligot} developed \sqligot{}, an SQL injection detector that represents a SQL query as a graph, both directed and undirected.
Each node in this graph is a token of the SQL language, plus all system reserved and user defined table names, variables, procedures, views and column names.
Moreover, the edges are weighted uniformly or proportionally to the distances in terms of adjacency.
We omit all the details of the model, as they are well described in the paper.
Kar et al. released the hyper-parameter they found on their dataset, but since both $C$ and $\gamma$ depend on data, we had to train these models from scratch.

We performed a 10-fold cross-validation for \sqligot{}, using \benignData{} benign and \injData{} malicious queries, again using the \scikit{} library.
After the feature extraction phase, the dataset is shrunk to: $(i)$ \sqligotDBenignData{} sane \sqligotDInjData{} and malicious data for the directed graph versions, $(ii)$ and \sqligotUBenignData{} sane \sqligotUInjData{} and malicious data for the undirected graph versions of \sqligot{}.
Again, many queries possess the same structure as others, and this is likely to happen for sane queries.
As already said in the previous paragraph, we are dealing with imbalance between the two classes, and we treat this issue by using a balanced accuracy loss function, provided by the \scikit{} framework.
Table~\ref{table:token_based} shows the result of the training phase of the different \sqligot{} classifiers.
Both the hyper-parameters and the scores are almost the same for all the different versions of \sqligot{}.


\subsection{Benchmark}
\label{subsec:benchmark}
We carried out benchmark experiments to assess the detection rates of the classifiers discussed above.
For all the classifiers used for this benchmark, we formed a dataset of 8,000 sane queries and 8,000 SQL injection queries, and we classified them using the models we have trained.
Table~\ref{table:benchmark} shows the results of our experiment.
\begin{table}[]
    \begin{tabular}{llrrr}
        \toprule
                                & & \textbf{A} & \textbf{R} & \textbf{P} \\
                                \midrule
        \multirow{3}{*}{\modsecurity{} CSR} & Paranoia 1/2 & 86.10\% & 86.10\% & 100\%\\
                                            & Paranoia 3/4 & 91.85\% & 91.85\% & 100\%\\
                                            & Paranoia 5   & 96.46\%  &96.46\% & 100\%\\
                                            \midrule
        \wafbrain{}  & RNN & 98.27\%  & 96.73 & 99.8\% \\\midrule
        \multirow{4}{*}{Token-based}
                     & Naive Bayes    & 50.16\% & 98.71\% & 50.08\%\\
                     & Random forest  & 98.33\% & 98.33\% & 100\%\\
                     & Linear SVM   & 98.75\% & 98.76\% & 100\%\\
                     & Gaussian SVM   & 97.82\% & 97.82\% & 100\%\\
                     \midrule
                     \multirow{4}{*}{\sqligot{}}
                     & Dir. Prop.    &90.61\%&97.30\%&85.82\%\\
                     & Undir. Prop.  & 96.38\% & 97.31\% & 95.54\%\\
                     & Dir. Unprop.  & 90.52\% & 97.12\% &85.80\% \\
                     & Undir. Unprop.& 96.25\% & 97.05\% &95.53\% \\
                     \bottomrule
    \end{tabular}
    \caption{Benchmark table.}\label{table:benchmark}
\end{table}

We evaluated the performance of each classifier by accounting three different metrics: $(i)$ \emph{accuracy}, $(ii)$ \emph{recall}, and $(iii)$ \emph{precision}.
We denote the true positives as $TP$, true negatives as $TN$, false positives as $FP$ and false negatives as $FN$.
Accuracy is computed as $A = \frac{TP + TN}{TP+TN+FP+FN}$, recall is computed as $R = \frac{TP}{TP + FN}$ and precision is computed as $P = \frac{TP}{TP + FP}$.
The accuracy measures how many samples have been correctly classified, i.e., a sane query classified as sane or an injected query classified as malicious.
The recall measures how good the classifier is at identifying samples from the relevant class, in this case the injection payloads.
Scoring a high recall value means that the classifier labeled most of the real positives in the dataset as positives.
The precision measures how many of the samples classified as relevant are actually relevant.

Since the Naive Bayes algorithm tries to discriminate between input classes by considering each variable independent one to another, it misses the real structure of the SQL syntax.
Hence, it cannot properly capture the complexity of the problem.
All other classifiers may be compared with different levels of paranoia offered by \modsecurity{}, showing their effectiveness as WAFs.
\wafbrain{} results are comparable to what the author claims on his GitHub repository.

\begin{figure*}
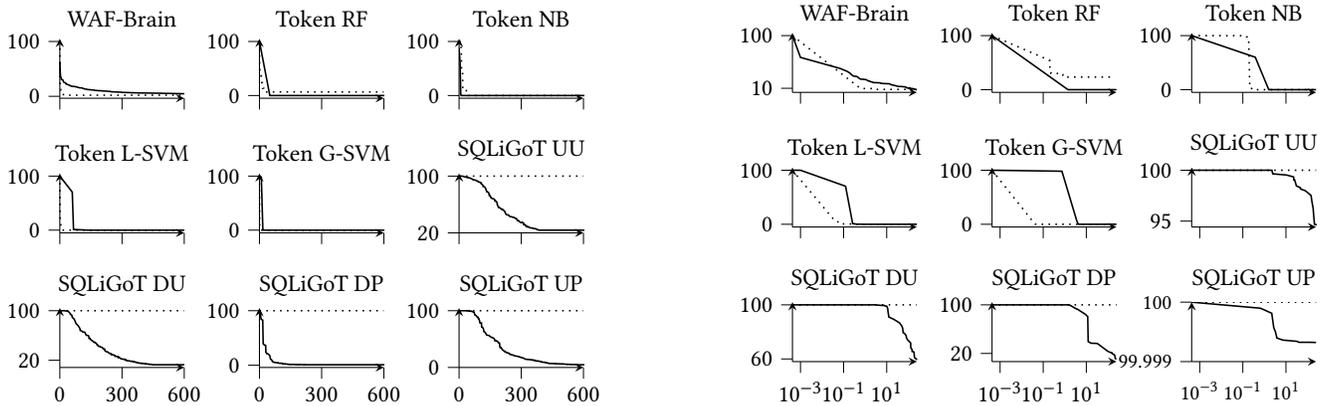

    \input{fig/eval_comparison_over_mutations}
    \hspace{40pt}
    \input{fig/eval_comparison_over_time}

    \caption{\emph{Guided} (solid) vs. \emph{unguided} (dotted) search strategies applied to initial payload \texttt{admin' OR 1=1\#}.}\label{fig:attack_plot}
\end{figure*}

\section{Evading Machine Learning WAFs}\label{sec:attacks}

In this section, we experimentally assess \toolname{} against the classifiers introduced above.
The experiments were performed on a DigitalOcean\footnote{\url{https://www.digitalocean.com/}} droplet VM with 6 CPUs and 16GB of RAM.
For a baseline comparison we used an \emph{unguided} mutational fuzzer.
The unguided fuzzer randomly applies the mutation operators of Section~\ref{sec:mutation}.
Moreover, we executed 100 instances of the unguided fuzzer on each classifier.
Then, we compared a single run of \toolname{} against the best payload generated by the 100 unguided instances over time.
Both the \toolname{} and the unguided fuzzers were configured to start from the payload \verb|admin' OR 1=1#|, initially detected with 100\% confidence by each classifier.

\subsection{Assessment results}\label{subsec:attack_result}

Figure~\ref{fig:attack_plot} shows the evolution of the confidence score for each classifier.
In each plot, we compare the best sample obtained by \toolname{} (solid line) and the best sample generated by all the 100 processes of the \emph{unguided} fuzzer (dashed line).

The first group of plots (left) show the evolution of the confidence scores against the number of mutation rounds.
The second group (right), shows the confidence score over the actual time of computation. 
In particular, we show the first 10 seconds of computation.
Since some scores quickly degrade in the first milliseconds of computation, we report the $x$ axis in log scale.

\subsection{Interpretation of the results}
\label{subsec:interpretation}

Our experiments highlight a few facts that we discuss below.

\paragraph{Feature choice matters}

As explained in Section~\ref{sec:our-wafs}, all the considered classifiers are based on syntactic features.
However, different feature set change the robustness of a classifier.
For instance, \wafbrain{} quickly lost confidence when the payload mutated, because 
\wafbrain{} is trained from uninterpreted, fixed-length sequences of characters and
our mutation operators can enlarge a payload beyond the adequacy of the length assumed by \wafbrain{}.
Also \tokenbased{} classifiers do not perform well against mutations.
The reason is that malicious and benign payloads overlap in the feature space.
All \sqligot{} versions showed to be robust against the unguided approach.
These classifiers use the SVM algorithm as some of the token based classifiers, but their feature set imposes more structure inside the feature representation.
Hence, random mutations have a negligible probability to evade them.
Instead, since \toolname{} relies on a guided strategy, it can effectively craft adversarial examples (although more effort is needed).

\paragraph{Finding adversarial examples is non-trivial}

\sqligot{} classifiers resist the unguided evaluation as it is unlikely that a mutation can move the sample away from a plateau region where the confidence of being a SQL injection is high.
The main reasons are: $(i)$ \sqligot{} considered a large number of tokens (so reducing the collision problem that affects other classifiers, since the compression factor applied by the feature extractor is lower);
$(ii)$ The structure of the feature vector is inherently redundant, i.e., each pair of adjacent variables describe the same token;
$(iii)$ the models are regularized, hence the decision function is smoother between input points and it manages to generalize over new samples.

\paragraph{\toolname{} effectively evades WAFs}
Moving randomly in the input space is not an effective strategy.
\toolname{} finds adversarial samples by leveraging on hints given by classifier outputs.
The guided approach accomplishes what the unguided approach failed to, by moving points away from plateaus and putting them in regions of low confidence of being recognized as SQL injection.
Moreover, among the \sqligot{} classifiers, the \textit{undirected unproportional} is the most resilient variant.
Recalling the definition of the algorithm~\cite{kar2016sqligot}, the feature extractor assigns uniform weights to tokens in the same window instead of balancing the score w.r.t. the distance of the current token. Hence, the classifier gains some invariance over the sequence of extracted tokens, making it more robust to adversarial noise.

\subsection{Discussion and limitations}
\label{sec:avoiding-adversarial}

Our experiments show that, starting from a target malicious payloads, \toolname{} effectively degrades the confidence scores of the considered classifiers.
In this section we discuss implications and limitations of this result.

\paragraph{Generality of the experiments}
As discussed in Section~\ref{sec:dataset}, the classifiers were trained with a dataset that we had to build from scratch.
This has clear consequences on our experimental results.
Hence, to extend the validity of our results, new experiments should be executed from other, real-world, datasets.
%

Another limitation is that we did not take into account the robustness of WAFs combining signatures \emph{and} ML techniques, called \emph{hybrid}.
These systems are becoming more and more common.

\paragraph{Adversarial attacks mitigation}
Demontis et al.~\cite{demontis2018intriguing} showed the effect of the presence of regularization when a classifier is under attack.
Without regularization an attacker may craft an adversarial example against the target, due to the high irregularity of the victim function.
Adding the regularization parameter has the effect of smoothing the decision boundary of the function between samples, reducing the amount of local minima and maxima.
On top of that, the adversary needs to increase the amount of perturbations to craft adversarial examples.
All models we trained have been properly regularized.

Grosse et al.~\cite{grosse2017statistical} propose the so called \emph{adversarial training}, that basically is a re-fit of the classifier also including the attack points.
This defense system leads to better robustness against adversarial examples, at the cost of worse accuracy scores.
Again, as shown by Carlini et al.~\cite{carlini2017adversarial} this is not a solution, but it may slow down the adversary in finding adversarial examples.

\section{Related work}

In this section, we present some work related to WAFs, as well as evasion techniques that have been proposed to bypass them.

\emph{Attacks against signature-based WAFs:}
Appelt et al.~\cite{appelt15behindwaf,appelt18mldriven} propose a technique to bypass signature-based WAFs.
Their technique is a search-based approach in which they create new payloads from existing blocked payloads.
The problem with implementing a search-based approach in this context is hard:
the obvious evaluation function for a payload against the target WAF is a decision
function with values \verb|PASSED/BLOCKED|.
Search-based approaches perform poorly if the evaluation function has many plateaus.
To mitigate this issue, the authors propose an approximate evaluation function which returns
the probability of a payload of being ``near'' the \verb|PASSED| or \verb|BLOCKED| state.
In the best case scenario, this function smooths the plateau and the search algorithm
converges to the \verb|PASSED| state.

\emph{Automata based WAFs:}
Halfond et al.~\cite{halfond2005amnesia} propose \emph{AMNESIA}, a tool to detect and prevent SQL injection attacks.
The algorithm works by creating a Non-Deterministic Finite Automa representing all the SQL queries that the application can generate.
The main issues with this approach are that an attack can bypass it ($i$) if the model is too conservative and includes queries that cannot be generated by the application or ($ii$) if the attack has the same structure of a query generated by the application.
Bandhakavi et al.~\cite{bandhakavi2007candid} developed \emph{CANDID}, a tool that detects SQL injection attempts via candidate selection.
This approach consists of transforming queries into a canonical form and evaluating each incoming query against candidate ones generated by the application.

\emph{Machine learning WAFs:}
Ceccato et al.~\cite{ceccato2016sofia} propose a clustering method for detecting SQL injection attacks against a victim service.
The algorithm learns from the queries that are processed inside the web application under analysis, using an unsupervised one-class learning approach, namely K-medoids~\cite{rdusseeun1987clustering}.
New samples are compared to the closest medoid and flagged as malicious if their edit distance w.r.t.\ the chosen medoid is higher than the diameter of the cluster.
Kar et al.~\cite{kar2016sqligot} develop \sqligot, a support vector machine classifier (SVM)~\cite{cortes1995support} that expresses queries as graphs of tokens, whose edges represent the adjacency of SQL-tokens.
This is the classifier we used in our analysis.
Pinzon et al.~\cite{pinzon2013idmas} explore two directions: visualization and detection, achieved by a multi-agent system called \emph{idMAS-SQL}.
To tackle the task of detecting SQL injection attacks, the authors set up two different classifiers, namely a Neural Network and an SVM\@.
Makiou et al.~\cite{makiou2014improving} developed an hybrid approach that uses both machine learning techniques and pattern matching against a known dataset of attacks.
The learning algorithm used for detecting injections is a Naive Bayes~\cite{maron1961automatic}.
They look for 45 different tokens inside the input query, chosen by domain experts.
Similarly, Joshi et al.~\cite{joshi2014sql} use a Naive Bayes classifier that,
given a SQL query as input, extracts syntactic tokens using spaces as separator.
The algorithm produces a feature vector that counts how many instances of a particular word occurs in the input query.
The vocabulary of all the possible observable tokens is set a priori.
Komiya et al.~\cite{komiya2011classification} propose a survey of different machine learning algorithms for SQL injection attack detection.

\emph{Evading machine learning classifiers:}
The techniques that are used in the \sota{} are divided in two different categories: $(i)$ gradient and $(ii)$ black-box methods.
For a comprehensive explanation of these techniques, Biggio et al.\cite{biggio2018wild} expose the \sota{} of adversarial machine learning in detail.
The attacker can compute the gradient of the victim classifier w.r.t. the input they use to test the classifier.
Biggio et al.\cite{biggio13-ecml} propose a technique for finding adversarial examples against both linear and non linear classifiers, by leveraging on the information given by the gradient of the target model.
Similarly, Goodfellow et al.\cite{goodfellow2014explaining} present Fast Gradient Sign Method (FGSM), which is used to perturb images to shift the confidence of the real class towards another one.
Papernot et al.~\cite{papernot2016limitations} propose an attack that computes the best two features to perturb in order to most increase the confidence of it belonging to a certain class. This method leverages on gradient information too.
If the attacker has information regarding a particular system, but they can not access it, they can try to learn a surrogate classifier, as proposed by Papernot et al.\cite{papernot2017practical}.
Many papers that craft attacks in other domains~\cite{demetrio2019explaining,rosenberg2018generic,kolosnjaji2018adversarial} belong to this category.
If the attacker does not have access to the model, or they have no information on how to reconstruct it locally, they treat this case as a black-box optimization problem.
Ilyas et al.~\cite{ilyas2018black} apply an evolution strategy to limit the number of queries that are sent to the victim model to craft an adversarial example.
Xu et al.~\cite{xu2016automatically} propose a technique that uses a genetic algorithm for crafting adversarial examples that bypass PDF malware classifiers. Anderson et al.~\cite{anderson2017evading} evade different malware detectors by altering malware samples using semantics invariant transformations, by leveraging only on the score provided by the victim classifier.


\section{Conclusion}\label{sec:conclusion}

We provided experimental evidence that machine learning based WAFs can be evaded.
Our technique takes advantage of an adversarial approach to craft malicious payloads that are classified as benign.
Moreover, we showed that \toolname{} efficiently converges to bypassing payloads.
We show the results of this technique applied to existing WAFs, both via a guided and unguided approach.
We leveraged on a set of syntactic mutations that do not alter the original semantics of the input query.
Finally, we built a dataset of SQL queries and we released it publicly.

Our work highlights that machine learning based WAFs are exposed to a concrete risk of being bypassed.
Future directions include testing WAFs based on other techniques such as hybrid ones,
finding new mutations to improve our approach,
and take advantage of our adversarial technique to improve detection of malicious payloads.


\bibliographystyle{ACM-Reference-Format}
\bibliography{reference,aml,wafaml}

\end{document}